\documentclass[12pt]{article}
\usepackage{amsmath}
\usepackage{graphicx,psfrag,epsf}
\usepackage{enumerate}

\usepackage[round]{natbib}
\usepackage{url} 
\usepackage{amsfonts}
\usepackage{amsthm}
\usepackage{mathrsfs}
\usepackage{multirow}
\usepackage{color}
\usepackage[titletoc,title]{appendix}
\usepackage[font={footnotesize}]{caption}
\usepackage[font={footnotesize}]{subcaption}
\usepackage{tabularx} 

\RequirePackage{hyperref}
\usepackage{longtable}

\makeatletter
\renewcommand{\author}[1]{\gdef\@author{\parbox{\textwidth}{\raggedright #1}}}
\makeatother

\newcommand{\vx}{\mathbf{x}}
\newcommand{\mX}{\mathbf{X}}
\newcommand{\vv}{\mathbf{v}}
\newcommand{\mV}{\mathbf{V}}
\newcommand{\mW}{\mathbf{W}}
\newcommand{\mA}{\mathbf{A}}
\newcommand{\vmu}{\boldsymbol{\mu}}
\newcommand{\mSigma}{\boldsymbol{\Sigma}}
\newcommand{\mS}{\mathbf{S}}
\newcommand{\mI}{\mathbf{I}}

\begin{document}
	
\bibliographystyle{unsrtnat}
	
\def\spacingset#1{\renewcommand{\baselinestretch}{#1}\small\normalsize} \spacingset{1}
	
\title{PSInference: A Package to Draw Inference for Released Plug-in Sampling Single Synthetic Dataset}
\author{
{\small Ricardo Moura$^{1,2}$\thanks{Corresponding author: rp.moura@fct.unl.pt}, Mina Norouzirad$^{1}$, V\'{i}tor Augusto$^{2}$ and Miguel Fonseca$^{2,3}$}\\[2em]
{\footnotesize $^{1}$Center for Mathematics and Applications (NOVA Math), NOVA School of Science and }\\ 
{\footnotesize Technology (NOVA FCT), NOVA University Lisbon, Caparica, Portugal}\\
{\footnotesize $^{2}$Portuguese Navy Research Center (CINAV), Naval Academy, 
Portuguese Navy, Alfeit, Portugal}\\
{\footnotesize $^{3}$Department of Mathematics, NOVA School of Science and Technology (NOVA FCT),}\\ 
{\footnotesize NOVA University Lisbon, Caparica, Portugal}
}
\date{}
\maketitle
	
\begin{abstract} 
The development and generation of synthetic data are becoming increasingly vital in the field of statistical disclosure control. The PSInference package provides tools to perform exact inferential analysis on singly imputed synthetic data generated through Plug-in Sampling assuming that the original dataset follows a multivariate normal distribution. Includes functions to test the synthetic data's covariance structure, covering aspects like generalized variance, sphericity, independence between subsets of variables, and regression of one set of variables on another. This package addresses the gap in the existing software by providing exact inferential methods suitable for cases where only a single synthetic dataset is released.\\

\end{abstract}
	
\spacingset{1.5}
	
\section{Introduction}\label{sec:1}

Statistical disclosure control (SDC) techniques are used to balance the dual objectives of releasing statistical information from surveys while protecting the confidentiality of respondents' data. To fulfill those two objectives, there are a range of methods, such as swapping data values, introducing random noise through addition or multiplication, and generating synthetic datasets to minimize the risk of disclosure \citep{hundepool2010handbook}.

The development and dissemination of synthetic data are becoming increasingly vital in the field of statistical disclosure control, where preserving the privacy of sensitive information is a primary concern \citep{klein2013comparison}. The main reason for this increase in attention is due to the fact that it maintains the statistical patterns of the assumed model for the original data, which is not the case for most of the other SDC techniques \citep{Drechsler2011}.

Most inferential approaches for synthetic data rely on the access to multiple imputation synthetic generated data \citep{Raghunathan2003,Reiter2003,Reiter2005c} and are based on the works of \citet{Rubin1987} and \citet{Little1993}. However in the last decade \citet{Klein2015a,Klein2015b,Klein2015c, Klein2015d}, \citet{Moura2016,Moura2017,Moura2021} and \citet{Klein2021} have developed a set of exact parametric inferential methods that are applicable to singly imputed synthetic data assuming a variety of probability models. The motivation relies on the fact that there are practical scenarios where only a single synthetic dataset \citep{abowd2020,Kinney2011,Kinney2014,Bowen2020,Alam2020} is released due to constraints such as computational resources, data management policies, or privacy considerations and that the traditional asymptotic methods for combining inferences present in the literature are only suitable to draw inference when multiple synthetic versions of the original data are available. It is also crucial to note that all of the work done in the field of single imputation by those authors who were previously mentioned can be extended to multiple imputation. These inferential procedures are not asymptotic but exact, making them more robust even for small sample sizes and even when only a limited number of synthetic versions are available.

Packages such as \textit{synthpop} \citep{Nowok2016} and \textit{mice} \citep{Buuren2011} were developed to generate synthetic data, but they do not include functions specifically designed to directly perform statistical inference on synthetic datasets. Instead, users must employ standard statistical methods after synthesis, adhering to combination rules like those proposed by \citet{Reiter2003}. In contrast, the existing software tools that support inference, such as \textit{mitools} \citep{mitools}, are primarily intended for application in multiple imputation settings and are not meant to handle the unique challenges posed by synthetic data generated for disclosure control purposes.

Building on this concept and looking to the work done in \citet{Klein2021}, our package, \href{https://cran.r-project.org/web/packages/PSinference/index.html}{PSInference}, provides a suite of tools for performing inference on singly imputed synthetic data generated via Plug-in sampling (PS) when the original data is assumed to come from a multivariate normal model. The core methodologies implemented in \href{https://cran.r-project.org/web/packages/PSinference/index.html}{PSInference} are grounded in the theoretical framework developed by \citet{Klein2021}. The package includes functions for testing various characteristics of the underlying population's covariance structure, such as the generalized variance, the sphericity, the independence between two subsets of variables, and the regression of one set of variables on another.

It also includes a function that enables a user to build a singly PS-based synthetic version of the original data, as long as the original data comes from a multivariate normal model. Even if not created for that purpose, this function can also be used to create multiple synthetic versions.

The package presented in this work aims to be part of a series of tools designed to support users working with synthetic datasets. We aim to establish a comprehensive suite of functions that will expand the available analysis options. We plan to expand beyond this first effort in future packages by integrating additional functions that make it possible for flexible and robust statistical analysis of synthetic data across various applications.

The paper is structured as follows: Section \ref{sec:Notation} describes the notation and outlines the key mathematical concepts and assumptions that serve as the foundation for this package. A detailed explanation and implementation of \href{https://cran.r-project.org/web/packages/PSinference/index.html}{PSInference} are provided in Section \ref{sec:3}, which also provides an overview of the characteristics of the package. Section \ref{sec:4} illustrates the practical applications of the package and provides code to assist users with implementation. Finally, Section \ref{sec:5} summarizes the key findings and contributions of the work, suggesting potential directions for future research.

\section{Notation and Models}\label{sec:Notation}

Before looking into the models that form the basis of our package, it is important to first introduce the concepts of fully synthetic data and partially synthetic data.

\subsection{Fully synthetic data}

Fully synthetic data refers to datasets where all original data instances are replaced with synthetic values generated from a statistical parametric model or from a non-parametric model. This approach aims to maximize privacy by ensuring that no original data records are directly disclosed while still preserving the statistical properties needed for valid analysis \citep{Moura2024}. 

\subsection{Partially synthetic data}

Partially synthetic data, by contrast, involves replacing only a subset of the original data, usually the sensitive values, with synthetic versions generated from a model (parametric or not). One can employ several strategies for this substitution: one can replace all deemed sensitive values in a respondent's record, replace all values for a selected subset of (sensitive) variables, or only replace specific values that could potentially reveal an individual's identity or sensitive information. This method enhances data utility while still preserving sensitive information, as non-sensitive values remain unchanged. However, it is clear that this improvement on the data quality can potentially compromise the ability to protect confidentiality \citep{Moura2024}.

\subsection{Generating fully synthetic data}

The theoretical foundation of this package is based on the assumption that the original dataset comes from a multivariate normal distribution. To this end, it is considered that the vector of variables is defined as \(\vx = \left(x_1, \dots, x_p\right)^\top\), where all variables are regarded as sensitive in this context. 
Therefore, the original (confidential) dataset will be defined as
\[
\mX=\left(\vx_1,\dots,\vx_n\right)=
\begin{bmatrix}
x_{11} & \dots & x_{1n}\\
\vdots & \ddots & \vdots \\
x_{p1} & \dots & x_{pn}
\end{bmatrix}.
\]
where, for all $i = 1, \ldots, n$, $\vx_i = (x_{1i}, \ldots, x_{pi})^{\top}$.

We assume that \(\mX\) will be normally distributed, \textit{i.e.}, 
\begin{equation}\label{eq:x_i}
    \vx_i \sim \mathcal{N}_p(\vmu, \mSigma), i=1,\dots,n,
\end{equation}
where \(\vmu\) is the mean vector, and \(\mSigma\) denotes the covariance matrix.

To generate a fully synthetic version of the original data, we will need the sample mean \(\bar{\vx} = \frac{1}{n} \sum_{i=1}^n \vx_i\) and the sample covariance matrix \(\hat{\mSigma} = \frac{1}{n-1}{\mS}\), where 
\(
\mS = \sum_{i=1}^n (\vx_i - \bar{\vx})(\vx_i - \bar{\vx})^\top
\)
is the sample Wishart matrix, where it is distributed as a Wishart distribution with \((n-1)\) degrees of freedom and covariance matrix \(\mSigma\).

The underlined idea of the PS Sampling method of generating data is that we plug-in the sample estimates in the original assumed model and randomly generate a new dataset. Specifically, we generate
\[
\mV=\left(\vv_1,\dots,\vv_n\right)=
\begin{bmatrix}
v_{11} & \dots & v_{1n}\\
\vdots & \ddots & \vdots \\
v_{p1} & \dots & v_{pn}
\end{bmatrix},
\]
where, for all $i = 1, \ldots, n$, $\vv_i = (v_{1i}, \ldots, v_{pi})^{\top}$,
by randomly drawing, independently,
\begin{equation}\label{eq:synmodel}
\vv_i \sim \mathcal{N}_p(\bar{\vx}, \widehat{\mSigma}), i=1,\dots,n.
\end{equation}

To generate more than one synthetic version of the original dataset, one simply needs to repeat the process \(M\) times, where \(M\) is the number of synthetic versions required.

To be perfectly clear, the original data \(\mX\) is the data that we consider to be fully confidential and that we may not share publicly, whereas the synthetic data \(\mV\) is the data that we may share publicly.

With access to \(\mV\), the released version of the dataset, one should note that \(\bar{\vv} = \frac{1}{n} \sum_{i=1}^n \vv_i\) is the maximum likelihood (also unbiased) estimator of \(\vmu\), that 
\begin{equation}\label{eq:Sigma_star}
    \mSigma^\star=\frac{1}{n-1}\mS^\star=\frac{1}{n-1}\sum_{i=1}^n (\vv_i - \bar{\vv})(\vv_i - \bar{\vv})^\top
\end{equation}
is an unbiased estimator of \(\mSigma\), and that the distribution of \(\mS^\star\), conditionally on \(\mS\) is \citep{Klein2021}
\begin{equation}\label{eq:S_star}
\mS^\star\sim \mathcal{W}_p\left(n-1,\frac{1}{n-1}\mS\right),
\end{equation}
where \(\mathcal{W}_p\left(\nu,\mA\right)\) is the Wishart distribution with  \(\nu\) degrees of freedom and  a \(p\times p\) scale matrix \(\mA\).

\subsection{Generalized variance}

Based on Eq. \eqref{eq:S_star}, \citet{Klein2021} concluded that
\begin{equation}\label{eq:T_1}
T_1^\star = (n-1)\frac{|\boldsymbol{S}^*|}{|\boldsymbol{\Sigma}|}
\end{equation}
would be stochastically equivalent to
\begin{equation}\label{eq:T_1dist}
    \left(\prod_{j=1}^p A_j\right) \left(\prod_{j=1}^p B_j\right)
\end{equation}
where \(A_j\) and \(B_j\), for \(j=1,\dots,p\), are all independently distributed as \(\chi^2_{n-j }\), following a chi-square distribution with \(n-j\) degrees of freedom.

From Eq. \eqref{eq:T_1}, we can construct the $(1-\alpha)$ level confidence interval for $|\mSigma|$ which will be 
\[
\left(\frac{(n-1)^p|\mS^\star|}{t^\star_{1,1-\alpha/2}}, \frac{(n-1)^p|\mS^\star|}{t^\star_{1,\alpha/2}} \right)
\]
where $t^\star_{1,\gamma}$ is the $\gamma$th percentile of $T_1$, obtained from \eqref{eq:T_1dist}.

\subsection{Sphericity test}

The assumption of sphericity structure of the covariance matrix is an important condition for the validity of many inferential tests, including multivariate analysis of variance (MANOVA), repeated measures ANOVA, and multivariate analysis of covariance (MANCOVA) \citep{Moura2024}. This assumption implies that the variances of the differences between all possible pairs of within-subject conditions are equal.

The sphericity test consists on testing the hypotheses
\[
\mathcal{H}_0: \mSigma = \sigma^2\mI_p~~ \text{vs.}~~\mathcal{H}_1:\mSigma\neq\sigma^2\mI_p.
\]

Considering that
\begin{equation}\label{eq:T_2}
T^\star_2=\frac{\left|\mS^\star\right|^{1/p}}{tr\left(\mS^\star\right)/p},
\end{equation}
\citet{Klein2021} demonstrated that its distribution, under \(\mathcal{H}_0\), is stochastically equivalent to
\begin{equation}\label{eq:T_2dist}
    \frac{\left|\mW_1\mW_2\right|^{1/p}}{\textrm{tr}(\mW_1\mW_2)/p} 
\end{equation}
where $\mW_1 \sim \mathcal{W}_p\left(n-1,(1\,/\,(n-1))\mI_p\right)$ and $\mW_2 \sim W_p\left(n-1,\mI_p\right)$, independent between each other.
From \eqref{eq:T_2dist} we may construct the empirical distribution of \(T_2^\star\) and reject the null hypothesis, for a level of significance  \(\alpha\), if the observed value of \(T_2^\star\) is less than \(t_{2;\alpha}^\star\), the \(\alpha\)th percentile of \eqref{eq:T_2dist}.

\subsubsection{Independence test}

Analyzing the independence between two subsets of variables is critical to understanding the variables' structure. It is applicable for feature selection, for dimensionality reduction, and various other processes. The independence test serves to determine whether two subsets of variables behave independently or if there are any statistical relations between the two subsets \citep{Moura2024}.

For the application of the independence test, $\mathbf{x}_i$ in \eqref{eq:x_i} must be divided into two subvectors or subsets of variables $\mathbf{x}_i=\begin{bmatrix}
    \vx_{1i}\\\vx_{2i}
\end{bmatrix}$ where \(\vx_{1i}\) will be of size $p_1$, and \(\vx_{2i}\) will be of size $p-p_1$. Consequently, \(\mSigma\) and \(\mS^\star\) will be partitioned as follows:
\begin{equation}\label{eq:partition}
\mSigma = \begin{bmatrix}
	\mSigma_{11} & \mSigma_{12} \\
	\mSigma_{21} & \mSigma_{22}
\end{bmatrix}, \quad \mS^\star = \begin{bmatrix}
	\mS^\star_{11} & \mS^\star_{12} \\
	\mS^\star_{21} & \mS^\star_{22}
\end{bmatrix},
\end{equation}
where both \(\mSigma_{11}\) and \(\mS^\star_{11}\) will be $p_1 \times p_1$ matrices.

To test the hypothesis of independence of the two subsets of variables, we test 
\begin{equation}\label{indepedence2}
    \mathcal{H}_0: \mSigma_{12}=\mathbf{0}~~ \text{vs.}~~\mathcal{H}_1:\mSigma_{12}\neq\mathbf{0}.
\end{equation}

\citet{Klein2021} used
\begin{equation}\label{eq:T_3}
    T_3^\star=\frac{\left|\mS^\star\right|}{\left|\mS^\star_{11}\right|\left|\mS^\star_{22}\right|},
\end{equation}
and showed that its distribution is stochastically equivalent to the distribution of
\begin{equation}\label{eq:T_3dist}
    \frac{\left|\boldsymbol{\Omega}_2\right|}{\left|\boldsymbol{\Omega}_{2,11}\right|\left|\boldsymbol{\Omega}_{2,22}\right|}
\end{equation}
where \(\boldsymbol{\Omega}_{2}\sim \mathcal{W}_p\left(n-1,\frac{1}{n-1}\boldsymbol{\Omega}_{1}\right)\), and where \(\boldsymbol{\Omega}_{1}\sim \mathcal{W}_p \left(n-1,\mI_p\right)\), considering that \(\boldsymbol{\Omega}_{2}\) is also partitioned as
\[
\begin{bmatrix}
		\boldsymbol{\Omega}_{2,11} & \boldsymbol{\Omega}_{2,12}\\
		\boldsymbol{\Omega}_{2,21} & \boldsymbol{\Omega}_{2,22}
	\end{bmatrix}\
\]
with \(\boldsymbol{\Omega}_{2,11}\) is a $p_1 \times p_1$ matrix.

For the independence test, \(\mathcal{H}_0\) should be rejected for a level of significance \(\alpha\), if the observed value of \(T_3^\star\) is less than \(t_{3;\alpha}^\star\), the \(\alpha\)th percentile of \eqref{eq:T_3dist}.

\subsubsection{Regression of One set of Variables on other}

To quantify the relationships between two different subsets of variables, one can perform a test for the regression of one set of variables on another. This test aims to determine how changes (variations in one subset of variables can predict/explain the variations in another \citep{Moura2024}. Specifically, the regression coefficients, represented by the matrix \(\boldsymbol{\Delta} = \mSigma_{12} \mSigma_{22}^{-1}\), consider the partition of \(\mSigma\) as in \eqref{eq:partition}, describe the conditional relationship between the two subsets of variables. These quantify how changes in the independent variables influence the dependent variables, assuming a multivariate normal framework. Moreover, this test will provide insights into the underlying structure and dependencies between the two subsets of variables.

The referred test consists of testing
 \begin{equation}\label{regression}
 \mathcal{H}_0: \boldsymbol{\Delta} = \boldsymbol{\Delta}_0 \quad \text{vs.} \quad \mathcal{H}_1: \boldsymbol{\Delta} \neq \boldsymbol{\Delta}_0,
 \end{equation}
 where \(\boldsymbol{\Delta}=\mSigma_{12}\mSigma_{22}^{-1}\), and where \(\boldsymbol{\Delta}_0\) is some specific \(p_1\times (p-p_1)\) matrix, considering $p_1\leq p_2$.
 
\citet{Klein2021} considered the test statistic
\begin{equation}\label{eq:T_4}
T_4^\star = \frac{\left| \left( \mS^\star_{12}\mS_{22}^{\star-1} - \boldsymbol{\Delta}_0 \right) \mS^\star_{22} \left( \mS^\star_{12}\mS_{22}^{\star-1} - \boldsymbol{\Delta}_0 \right)^\top \right|}{\left| \mS^\star_{11} - \mS^\star_{12}\mS_{22}^{\star-1}\mS^\star_{21} \right|},
\end{equation}
proving that \eqref{eq:T_4} is stochastically equivalent to
\begin{equation}\label{eq:T_4dist}    \frac{\left|\boldsymbol{\Omega}_{2,12}\boldsymbol{\Omega}_{2,22}^{-1}\boldsymbol{\Omega}_{2,21}\right|}{\left|\boldsymbol{\Omega}_{2,11}-\boldsymbol{\Omega}_{2,12}\boldsymbol{\Omega}_{2,22}^{-1}\boldsymbol{\Omega}_{2,21}\right|}.    
\end{equation}
where \(\boldsymbol{\Omega}_{2}\sim W_p\left(n-1,\frac{1}{n-1}\boldsymbol{\Omega}_{1}\right)\), and where \(\boldsymbol{\Omega}_{1}\sim W_p \left(n-1,\mI_p\right)\).

For the regression test, \(\mathcal{H}_0\) should be rejected at a level of significance  \(\alpha\) if the observed value of \(T_4^\star\) is greater than \(t_{4;\alpha}^\star\), the \(\alpha\)th percentile of \eqref{eq:T_4dist}.

\section{Description and Implementation of PSInference}\label{sec:3}

The \href{https://cran.r-project.org/web/packages/PSinference/index.html}{PSinference} package is an R package designed to generate singly imputed fully synthetic data using PS under the assumption that the original data follows a multivariate normal distribution. The package provides tools for performing various inferential procedures on the synthetic data, including tests for generalized variance, sphericity, independence between two subsets of variables, and regression of one set of variables on another. These procedures are implemented according to the methodologies described in Section \ref{sec:Notation}, ensuring robust and accurate statistical inference for synthetic datasets intended for release. It is freely available at the Comprehensive R Archive Network (CRAN) at \url{https://cran.r-project.org/web/packages/PSinference/index.html}.

The R package \href{https://cran.r-project.org/web/packages/PSinference/index.html}{PSInference} contains the following functions: \texttt{Canodist} (described in \eqref{eq:synmodel}), \texttt{GVdist} (described in \eqref{eq:T_1} using \eqref{eq:T_1dist}), \texttt{Inddist} (described in \eqref{eq:T_3} using \eqref{eq:T_3dist}), \texttt{Sphdist} (described in \eqref{eq:T_2} using \eqref{eq:T_2dist}), \texttt{partition}, and \texttt{SimSynthData}, described summarily in Table \ref{tab:table_1}. 

\begin{longtable}{p{3cm} p{2.5cm} p{4.6cm} p{3.5cm}}
\caption{A description of functions in the \href{https://cran.r-project.org/web/packages/PSinference/index.html}{PSInference} package, their arguments, and return values}
    \label{tab:table_1} \\

\hline 
\textbf{Function} & \textbf{Arguments} & \textbf{Description} & \textbf{Return} \\
\hline 
\endfirsthead

\multicolumn{4}{c}{\tablename\ \thetable\ -- \textit{A description of functions in the \href{https://cran.r-project.org/web/packages/PSinference/index.html}{PSInference} package, ... (Continued)}} \\[0.5ex]
\hline 
\textbf{Function} & \textbf{Arguments} & \textbf{Description} & \textbf{Return} \\
\hline 
\endhead

\hline \multicolumn{4}{r}{\textit{Continued on next page}} \\[0.5ex]
\endfoot

\hline 
\endlastfoot

\texttt{simSynthData()} & \texttt{X} & Matrix or dataframe \\ 
    & \texttt{n\_imp} & Sample size & Matrix of generated synthetic dataset \\
    \hline 
\texttt{GVdist()} & \texttt{nsample} & Sample size & \\
& \texttt{pvariates} & Number of variables & \\ 
& \texttt{iterations} & Number of iterations for simulating values from the distribution and finding the quantiles. Default is 10000 & Vector of simulated distribution values for generalized variance quantiles \\
\hline 
\texttt{Sphdist()} & \texttt{nsample} & Sample size & \\ 
& \texttt{pvariates} & Number of variables &\\
& \texttt{iterations} & Number of iterations for simulating values from the distribution and finding the quantiles. Default is 10000 & Vector of simulated distribution values for sphericity quantiles \\
\hline 
\texttt{Inddist()} & \texttt{part} & Length of the first subset of variables (\(p_1\)) & \\ 
& \texttt{nsample} & Sample size & \\
& \texttt{pvariates} & Number of variables & \\ 
& \texttt{iterations} & Number of iterations for simulating values from the distribution and finding the quantiles. Default is 10000 & Vector of simulated distribution values for independence test quantiles \\
\hline 
\texttt{canodist()} & \texttt{part} & Length of the first subset of variables (\(p_1\)) & \\ 
& \texttt{nsample} & Sample size & \\
& \texttt{pvariates} & Number of variables & \\ 
& \texttt{iterations} & Number of iterations for simulating values from the distribution and finding the quantiles. Default is 10000 & Vector of simulated distribution values for regression test quantiles \\
\hline 
\texttt{partition()} & \texttt{Matrix} & A matrix to split & \\
& \texttt{nrows} & Positive integer indicating the number of row blocks & \\
& \texttt{ncols} & Positive integer indicating the number of column blocks & List of partitioned sub-matrices \\
\hline 
\end{longtable}

We must consider that the functions \texttt{Inddist} and \texttt{Sphdist} require the number of variables belonging to the first subset to be specified. For this, the user needs to understand that the variables will be selected in the order they appear in the dataset.

\section{Numerical Studies and Demonstration of Programming Code}\label{sec:4}

Even though the entire theoretical background can be found in \citet{Klein2021}, where all the proofs are analytical, there remains a need for practical verification of simulated data.
We will use the practical verification of \citet{Klein2021} theory to illustrate the use of the \href{https://cran.r-project.org/web/packages/PSinference/index.html}{PSinference} package for generating synthetic data and conducting the inferential tests discussed in this paper.

For that purpose, under the multivariate normal model we will consider $p=4$, the number of variables, and $\vmu=(1,2,3,4)^\top$, the population mean. After, we will consider four
different scenarios for the four different tests, considering the covariance matrices
\begin{equation}
\label{cov_matrices}
    \mSigma_1=\mI_4, \quad \mSigma_2=5\mI_4,\quad \mSigma_3 = 
    \left(
    \begin{array}{cccc}
        1 & 0.5 & 0.5 & 0.5 \\
        0.5 & 1 & 0.5 & 0.5 \\
        0.5 & 0.5 & 1 & 0.5 \\
        0.5 & 0.5 & 0.5 & 1
    \end{array}
    \right), \quad \mSigma_4 = 
    \left(
    \begin{array}{cccc}
        1 & 0.5 & 0 & 0 \\
        0.5 & 2 & 0 & 0 \\
        0 & 0 & 3 & 0.2 \\
        0 & 0 & 0.2 & 4
    \end{array}
    \right)
\end{equation}
chosen to illustrate the performance for all tests presented in this work. We used Monte Carlo simulations with $10^5$ iterations and estimated the coverage probability ($cov$) for $\alpha=0.05$ significance level. The estimated coverage probability is the proportion of iterations where the observed values of the test statistics fall inside the non-rejection region. For every iteration the PS single imputed dataset is created using the models described in Section \ref{sec:Notation} and the sample sizes considered were $n=10, 20, 100, 500$.

\subsection{Average coverage probability of the Generalized Variance}

For the generalized variance, the illustrative example is applied to the covariance matrices \(\mSigma_3\) and \(\mSigma_4\) defined in \eqref{cov_matrices}. These matrices, along with the defined mean vector \(\vmu\), are used in each iteration to generate an original dataset, which is then used to create a synthetic dataset via the \texttt{simSynthData} function. In each iteration, the observed values of \(T_1^\star\) are calculated for both scenarios and stored to compute the coverage probabilities (\(cov\)) for the two cases considered. Finally, the two \(cov\) values are computed after calculating the quantiles for \(T_1^\star\) using the distribution obtained with \texttt{GVdist}, and the results are presented.

\begin{verbatim}
library(MASS)
library(PSinference)

# Set random seed for reproducibility
set.seed(123)

# Total number of Monte Carlo simulations
N <- 100000

# Sample size
n <- 10 #replace by 20, 100, 500 for all other cases

# Population mean vector
mu <- c(1, 2, 3, 4)

# Number of covariates (dimensionality)
p <- length(mu)

# Covariance matrix 1 (Diagonal matrix with variance 1)
sigma_1 <- diag(c(1, 1, 1, 1), 4, 4)
# Covariance matrix 2 (Diagonal matrix with variance 5)
sigma_2 <- diag(c(5, 5, 5, 5), 4, 4)

# Covariance matrix 3 (homogeneous)
sigma_3 <- matrix(c(1, .5, .5, .5,
                    .5, 1, .5, .5,
                    .5, .5, 1, .5,
                    .5, .5, .5, 1), 4, 4)

# Covariance matrix 4 (heterogeneous)
sigma_4 <- matrix(c(1, .5, 0, 0,
                    .5, 2, 0, 0,
                    0, 0, 3, .2,
                    0, 0, .2, 4), 4, 4)

# Get the generalized variance distribution from PSinference
Ts <- GVdist(nsample = n, pvariates = p, iterations = N)

# Quantiles for the simulated distribution at alpha = 0.05
q975 <- quantile(Ts, probs = c(.975))
q025 <- quantile(Ts, probs = c(.025))

# Initialize result vectors for two test datasets
T1 <- c()
T2 <- c()

# Monte Carlo loop for generating synthetic data and perform simulation
for (i in 1:N) {  
  # Generate original data samples from multivariate normal distribution 
  # for mu and sigma_3 and sigma_4
  x1 <- mvrnorm(n, mu, sigma_3)
  x2 <- mvrnorm(n, mu, sigma_4)
  
  # Generate PS synthetic data for both datasets using simSynthData 
  # from PSinference
  v1 <- simSynthData(x1)
  v2 <- simSynthData(x2)
  
  # PS estimates of mu (mean vector) for both datasets
  mean_v1 <- apply(v1, 2, mean)
  mean_v2 <- apply(v2, 2, mean)
  
  # Create matrices of mean values for vectorized subtraction
  MEANv1 <- matrix(mean_v1, n, 4, byrow = TRUE)
  MEANv2 <- matrix(mean_v2, n, 4, byrow = TRUE)
  
  # Compute the covariance matrices of the synthetic data
  # (could be done just using var function)
  s_star1 <- t(v1 - MEANv1) \%*\% (v1 - MEANv1)
  s_star2 <- t(v2 - MEANv2) \%*\% (v2 - MEANv2)
  
  # Calculate observed T*_1 statistics for both datasets
  T1temp <- ((n - 1)^p) * det(s_star1) / det(sigma_3)
  T2temp <- ((n - 1)^p) * det(s_star2) / det(sigma_4)
  
  T1[i] <- T1temp
  T2[i] <- T2temp
}

# Print the quantiles of the simulated T1 values (observed and distribution)
print(quantile(T1, probs = c(0, 0.1, .5, .9, 1)))
print(quantile(T2, probs = c(0, 0.1, .5, .9, 1)))
print(quantile(Ts, probs = c(0, 0.1, .5, .9, 1)))

# Compute average coverage probabilities

# Rejection rate for 1st observed T*_1
rej1 <- mean(T1 < q025) + mean(T1 > q975) 

# Rejection rate for 2nd observed T*_1
rej2 <- mean(T2 < q025) + mean(T2 > q975)  

cov1 <- 1 - rej1  # Coverage probability for 1st case
cov2 <- 1 - rej2  # Coverage probability for 2nd case

# Print the coverage probabilities
print(c(cov1, cov2))
\end{verbatim}

\subsection{Average coverage probability for the Sphericity Test}

Similar to the previous example/numerical studies,
we illustrate the Sphericity Scenario, but now using matrices \(\mSigma_1\) and \(\mSigma_2\) defined in \eqref{cov_matrices}, which are spherical matrices. Obviously, now we base our simulations on \(T_2^\star\) and the function \texttt{Sphdist}.

\begin{verbatim}
# Using the same given parameters, vectors and matrices

set.seed(123)  # Set random seed for reproducibility

# Monte Carlo simulation to generate distribution for T*_(2, alpha)
Ts <- Sphdist(n, p, N)

# Quantile alpha = 0.05 of the simulated values
q05 <- quantile(Ts, probs = c(.05))

# Initialize result vectors for the test statistics
T1 <- c()
T2 <- c()

# Monte Carlo loop to calculate T*_2 statistics for 
# two different covariance matrices
for (i in 1:N) {
  
  # Generate original data samples from multivariate normal distribution 
  # for mu and sigma_1 and sigma_2
  x1 <- mvrnorm(n, mu, sigma_1)
  x2 <- mvrnorm(n, mu, sigma_2)
  
  # Generate PS synthetic data for both datasets using simSynthData 
  # from PSinference
  v1 <- simSynthData(x1)
  v2 <- simSynthData(x2)
  
  # PS estimates of mu for both datasets
  mean_v1 <- apply(v1, 2, mean)
  mean_v2 <- apply(v2, 2, mean)
  
  # Create matrices of mean values for vectorized subtraction
  MEANv1 <- matrix(mean_v1, n, 4, byrow = TRUE)
  MEANv2 <- matrix(mean_v2, n, 4, byrow = TRUE)
  
  # Covariance matrices of the synthetic data
  s_star1 <- t(v1 - MEANv1) \%*\% (v1 - MEANv1)
  s_star2 <- t(v2 - MEANv2) \%*\% (v2 - MEANv2)
  
  # Calculate T*_2 statistics for both datasets
  T1temp <- (det(s_star1)^(1/p)) / sum(diag(s_star1))
  T2temp <- (det(s_star2)^(1/p)) / sum(diag(s_star2))
  
  T1[i] <- T1temp  # Store the statistic for dataset 1
  T2[i] <- T2temp  # Store the statistic for dataset 2
}

# Print quantiles of the T1 and T2 distributions
print(quantile(T1, probs = c(0, 0.1, 0.5, 0.9, 1)))
print(quantile(T2, probs = c(0, 0.1, 0.5, 0.9, 1)))
print(quantile(Ts, probs = c(0, 0.1, 0.5, 0.9, 1)))

# Calculate and print coverage probabilities
cov1 <- mean(T1 > q05)
cov2 <- mean(T2 > q05)
print(c(cov1, cov2))
\end{verbatim}

\subsection{Average coverage probability for the Independence Test}

The simulation process here follows the same structure of the previous two scenarios, applied to matrices \(\mSigma_1\) and \(\mSigma_4\) defined in \eqref{cov_matrices}. However, since the partition of \(\mathbf{S}^\star\) is needed, the \texttt{partition} function is used, as described in \eqref{eq:partition}, making \(p_1=1\) when using \(\mSigma_1\) and \(p_1=2\) when using \(\mSigma_4\). In this case, the numerical study uses the \texttt{Inddist} function to obtain the distribution of \(T_3^\star\).

\begin{verbatim}
# Using the same given parameters, vectors and matrices

set.seed(123)  # Set random seed for reproducibility

# partition sizes
p1 <- 1   # Partition size 1
p2 <- 2   # Partition size 2

# Perform Monte Carlo simulations to generate the distribution 
# for the test statistics
T1 <- Inddist(p1, n, p, N)  # Simulate for the first partition
T2 <- Inddist(p2, n, p, N)  # Simulate for the second partition

# Quantile alpha = 0.05 for the simulated values
q05_1 <- quantile(T1, probs = c(.05))
q05_2 <- quantile(T2, probs = c(.05))

# Initialize result vectors for the second set of test statistics
T1_1 <- c()
T2_1 <- c()

# Monte Carlo loop to calculate the T*_3 statistics for 
# different covariance matrices
for (i in 1:N) {
  
  # Generate original data samples from multivariate normal distribution 
  # with given mu and sigma_1 and sigma_4
  x1 <- mvrnorm(n, mu, sigma_1)  # Sample using sigma_1
  x2 <- mvrnorm(n, mu, sigma_4)  # Sample using sigma_4
  
  # Generate PS synthetic single data for both datasets
  v1 <- simSynthData(x1)
  v2 <- simSynthData(x2)
  
  # PLS estimates of mu for both datasets
  mean_v1 <- apply(v1, 2, mean)
  mean_v2 <- apply(v2, 2, mean)
  
  # Create matrices of mean values for vectorized subtraction
  MEANv1 <- matrix(mean_v1, n, 4, byrow = TRUE)
  MEANv2 <- matrix(mean_v2, n, 4, byrow = TRUE)
  
  # Compute the covariance matrices of the synthetic data
  s_star1 <- t(v1 - MEANv1) \%*\% (v1 - MEANv1)
  s_star2 <- t(v2 - MEANv2) \%*\% (v2 - MEANv2)
  
  # Partition the covariance matrices into subsets
  s_star1_11 <- as.matrix(partition(s_star1, p1, p1)[[1]])
  s_star1_22 <- as.matrix(partition(s_star1, p1, p1)[[4]])
  
  s_star2_11 <- partition(s_star2, p2, p2)[[1]]
  s_star2_22 <- partition(s_star2, p2, p2)[[4]]
  
  # Calculate the T*_3 statistics for both datasets

  # T*_3 for first dataset
  T1temp <- det(s_star1) / (det(s_star1_11) * det(s_star1_22)) 

  # T*_3 for second dataset
  T2temp <- det(s_star2) / (det(s_star2_11) * det(s_star2_22))  
  
  # Store the results in T1_1 and T2_1
  T1_1[i] <- T1temp
  T2_1[i] <- T2temp
}

# Print quantiles of the distributions for T1, T1_1, T2, and T2_1
print(quantile(T1, probs = c(0, 0.1, .5, .9, 1)))
print(quantile(T1_1, probs = c(0, 0.1, .5, .9, 1)))
print(quantile(T2, probs = c(0, 0.1, .5, .9, 1)))
print(quantile(T2_1, probs = c(0, 0.1, .5, .9, 1)))

# Average coverage calculations
cov1 <- mean(T1_1 > q05_1)  # Coverage for T1_1
cov2 <- mean(T2_1 > q05_2)  # Coverage for T2_1

# Print coverage probabilities
print(c(cov1, cov2))
\end{verbatim}

\subsection{Average coverage probability for the Regression Test}

For the regression test \(cov\) computation, we use the covariance matrices \(\boldsymbol{\Sigma}_3\) and \(\boldsymbol{\Sigma}_4\), as defined in \eqref{cov_matrices}. At each iteration, the observed values of the test statistic \(T_4^\star\) are computed for both scenarios, making \(p_1=2\) when using \(\mSigma_3\) and \(p_1=1\) when using \(\mSigma_4\). Here, the test statistic \(T_4^\star\) is based on the matrix of regression coefficients \(\boldsymbol{\Delta} = \boldsymbol{\Sigma}_{12} \boldsymbol{\Sigma}_{22}^{-1}\), with corresponding values of \(p_1\). The observed values of \(T_4^\star\) are stored to compute the coverage probabilities (\(cov\)) for the regression test and using the distribution obtained via the \texttt{Regdist} function, the corresponding \(cov\) values are computed and the results are presented.

\begin{verbatim}
# Using the same given parameters, vectors and matrices

set.seed(123)  # Set random seed for reproducibility
N <- 100000    # Number of Monte Carlo simulations

# Sample size and partition sizes
n <- 10  # Total sample size
p1 <- 2   # Partition size for the first set of variables
p2 <- 1   # Partition size for the second set of variables

# Partition Sigma_3 into submatrices
sigma3_11 <- as.matrix(partition(sigma_3, p1, p1)[[1]])  # Submatrix Sigma_11
sigma3_22 <- as.matrix(partition(sigma_3, p1, p1)[[4]])  # Submatrix Sigma_22
sigma3_12 <- as.matrix(partition(sigma_3, p1, p1)[[2]])  # Submatrix Sigma_12
sigma3_21 <- as.matrix(partition(sigma_3, p1, p1)[[3]])  # Submatrix Sigma_21

Delta1 <- sigma3_12 \%*\% solve(sigma3_22)  # Compute Delta_1 for later use

# Partition Sigma_4 into submatrices
sigma4_11 <- as.matrix(partition(sigma_4, p2, p2)[[1]])  # Submatrix Sigma_11
sigma4_22 <- as.matrix(partition(sigma_4, p2, p2)[[4]])  # Submatrix Sigma_22
sigma4_12 <- as.matrix(partition(sigma_4, p2, p2)[[2]])  # Submatrix Sigma_12
sigma4_21 <- as.matrix(partition(sigma_4, p2, p2)[[3]])  # Submatrix Sigma_21

Delta2 <- t(sigma4_12) \%*\% solve(sigma4_22)  # Compute Delta_2 for later use

# Simulate canonical distances (T1 and T2) with the initial partition values
T1 <- canodist(p1, n, p, N)
T2 <- canodist(p2, n, p, N)

# Quantiles for alpha = 0.05 of the simulated values (for coverage analysis)
q95_1 <- quantile(T1, probs = c(.95))
q95_2 <- quantile(T2, probs = c(.95))

# Initialize result vectors for storing the T*_4 statistics
T1_1 <- c()
T2_1 <- c()

# Monte Carlo loop to generate synthetic data and calculate T*_4 statistics
for (i in 1:N) {
  
  # Generate original data samples from multivariate normal distributions 
  # with mu and Sigma
  x1 <- mvrnorm(n, mu, sigma_3)  # Data generated with Sigma_3
  x2 <- mvrnorm(n, mu, sigma_4)  # Data generated with Sigma_4
  
  # Generate PLS synthetic data from the original samples
  v1 <- simSynthData(x1)
  v2 <- simSynthData(x2)
  
  # Compute the means for the synthetic datasets
  mean_v1 <- apply(v1, 2, mean)
  mean_v2 <- apply(v2, 2, mean)
  
  # Create matrices of mean values for later calculations
  MEANv1 <- matrix(mean_v1, n, 4, byrow = TRUE)
  MEANv2 <- matrix(mean_v2, n, 4, byrow = TRUE)
  
  # Covariance matrices of the synthetic data
  s_star1 <- t(v1 - MEANv1) \%*\% (v1 - MEANv1)
  s_star2 <- t(v2 - MEANv2) \%*\% (v2 - MEANv2)
  
  # Partition the synthetic covariance matrices
  s_star1_11 <- as.matrix(partition(s_star1, p1, p1)[[1]])
  s_star1_22 <- partition(s_star1, p1, p1)[[4]]
  s_star1_12 <- partition(s_star1, p1, p1)[[2]]
  s_star1_21 <- partition(s_star1, p1, p1)[[3]]
  
  s_star2_11 <- partition(s_star2, p2, p2)[[1]]
  s_star2_22 <- partition(s_star2, p2, p2)[[4]]
  s_star2_12 <- partition(s_star2, p2, p2)[[2]]
  s_star2_21 <- partition(s_star2, p2, p2)[[3]]
  
  # Calculate adjusted submatrices and Delta_star for T*_4 statistics
  s_star1_112 <- s_star1_11 - 
                 s_star1_12 \%*\% (solve(s_star1_22) \%*\% s_star1_21)
  Delta_star1 <- s_star1_12 \%*\% solve(s_star1_22)
  
  s_star2_112 <- s_star2_11 - 
                 s_star2_12 \%*\% (solve(s_star2_22) \%*\% s_star2_21)
  Delta_star2 <- s_star2_12 \%*\% solve(s_star2_22)
  
  # Calculate T*_4 statistics for both datasets
  T1temp <- det((Delta_star1 - Delta1) \%*\%
            (s_star1_22 \%*\% t(Delta_star1 - Delta1))) / det(s_star1_112)
  T2temp <- det((Delta_star2 - Delta2) \%*\% 
            (s_star2_22 \%*\% t(Delta_star2 - Delta2))) / det(s_star2_112)
  
  # Store the results in T1_1 and T2_1
  T1_1[i] <- T1temp
  T2_1[i] <- T2temp
}

# Print the quantiles for T1, T1_1, T2, and T2_1 distributions
print(quantile(T1, probs = c(0, 0.1, .5, .9, 1)))
print(quantile(T1_1, probs = c(0, 0.1, .5, .9, 1)))
print(quantile(T2, probs = c(0, 0.1, .5, .9, 1)))
print(quantile(T2_1, probs = c(0, 0.1, .5, .9, 1)))

# Calculate average coverage for T1_1 and T2_1 compared to the 95th  
# percentile of T1 and T2
cov1 <- mean(T1_1 < q95_1)
cov2 <- mean(T2_1 < q95_2)

# Print the coverage probabilities
print(c(cov1, cov2))

\end{verbatim}

To summarize the results found in the simulations, Table \ref{tab:cov} shows, for values of $n$ and selected covariance matrices defined in (\ref{cov_matrices}), the estimated coverage values for:
\begin{itemize}
    \item the confidence interval for the Generalized Variance under the column \textit{Gener. Variance} and selected $\mSigma_3$ and $\mSigma_4$;
    \item the Sphericity test under the column \textit{Sphericity} and selected $\mSigma_1$ and $\mSigma_2$;
    \item the Independence test under the column \textit{Independence} and selected $\mSigma_1$ and $\mSigma_4$, for $p_1=1$ and $p_1=2$, respectively;
    \item the test for the regression of one set of variables on the other under the column \textit{Regression} and selected $\mSigma_3$ and $\mSigma_4$, for $p_1=2$ and $p_1=1$, respectively.
\end{itemize}

\begin{table}[h!]
    \centering
    \begin{tabular}{c|cc|cc|cc|cc|}
         \multicolumn{1}{c}{} & \multicolumn{2}{c}{\textit{Gener. Variance}} & \multicolumn{2}{c}{\textit{Sphericity}} & \multicolumn{2}{c}{\textit{Independence}} & \multicolumn{2}{c}{\textit{Regression}} \\
         & $\mSigma_3$ & $\mSigma_4$ & $\mSigma_1$ & $\mSigma_2$ & $\mSigma_1$ & $\mSigma_4$ & $\mSigma_3$ & $\mSigma_4$\\
         n & & & & & $p_1=1$ & $p_1=2$ & $p_1=2$ & $p_1=1$\\
         \hline
         10 & 0.948 & 0.950 & 0.951 & 0.952 & 0.951 & 0.948 & 0.949 & 0.950\\
         20 & 0.949 & 0.949 & 0.949 & 0.949 & 0.949 & 0.950 & 0.949 & 0.950\\
         100 & 0.951 & 0.948 & 0.951 & 0.950 & 0.949 & 0.948 & 0.950 & 0.952\\
         500 & 0.951 & 0.948 & 0.950 & 0.951 & 0.949 & 0.948 & 0.951 & 0.950
    \end{tabular}
    \caption{Estimated coverage probability for the tests of section \ref{sec:Notation} for $n=10, 20, 100, 500,\ p_1=1,2,\ \vmu=(1,2,3,4)'$ and $\mSigma_1$, $\mSigma_2$, $\mSigma_3$ and $\mSigma_4$ defined in (\ref{cov_matrices}).}
    \label{tab:cov}
\end{table}

\begin{figure}[!ht]
\centering
\begin{subfigure}{0.50\textwidth}
    \centering
    \includegraphics[width=\textwidth]{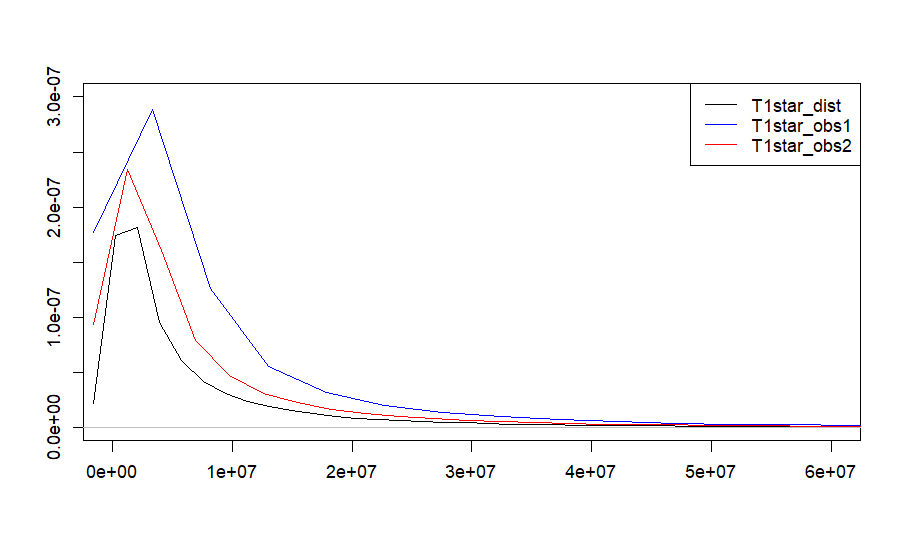}
    \caption{$n=10$}
\end{subfigure}
\begin{subfigure}{0.44\textwidth}
    \centering
    \includegraphics[width=\textwidth]{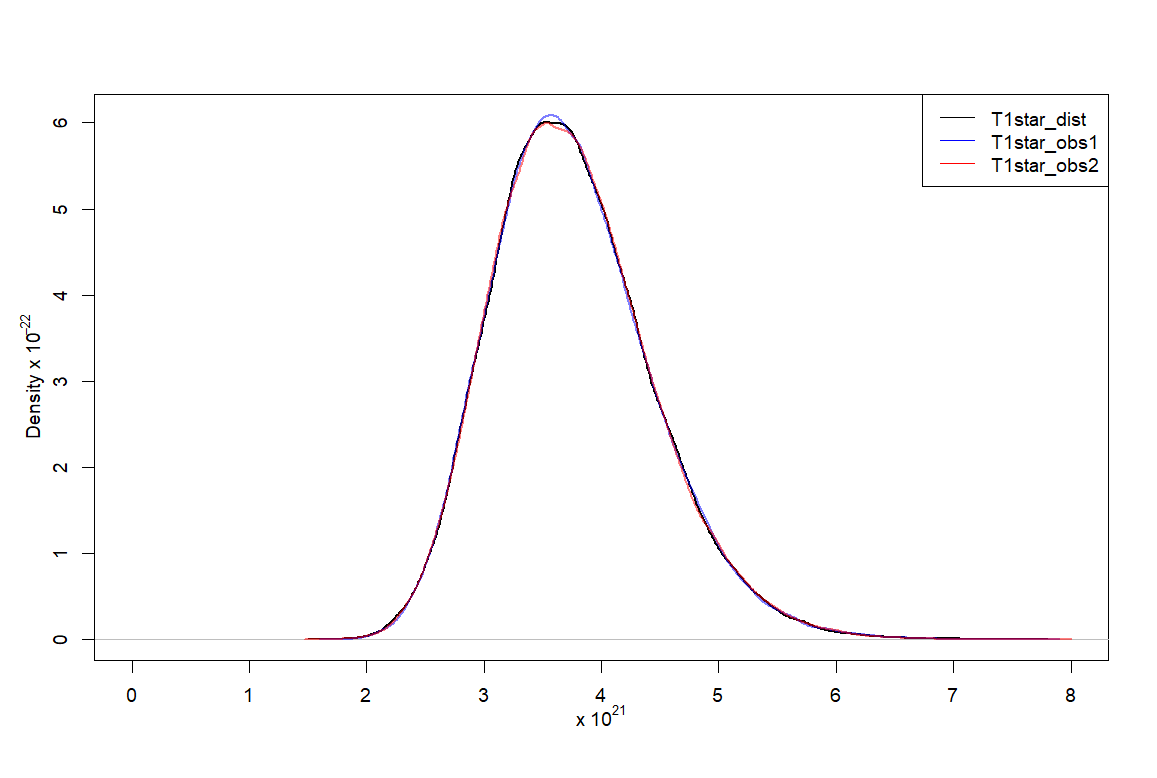}
    \caption{$n=500$}
\end{subfigure}
\caption{Graph for the Generalized Variance distribution along with the empirical distribution of the observed values of $T_1^\star$ for given $n=10, 500$, $p=4$, \(\vmu\), \(\mSigma_3\) (T1star\_obs1) and \(\mSigma_4\) (T1star\_obs2).}
\label{fig:GV}
\end{figure}

\begin{figure}[!ht]
\begin{subfigure}{0.48\textwidth}
  \centering
  \includegraphics[width=\linewidth]{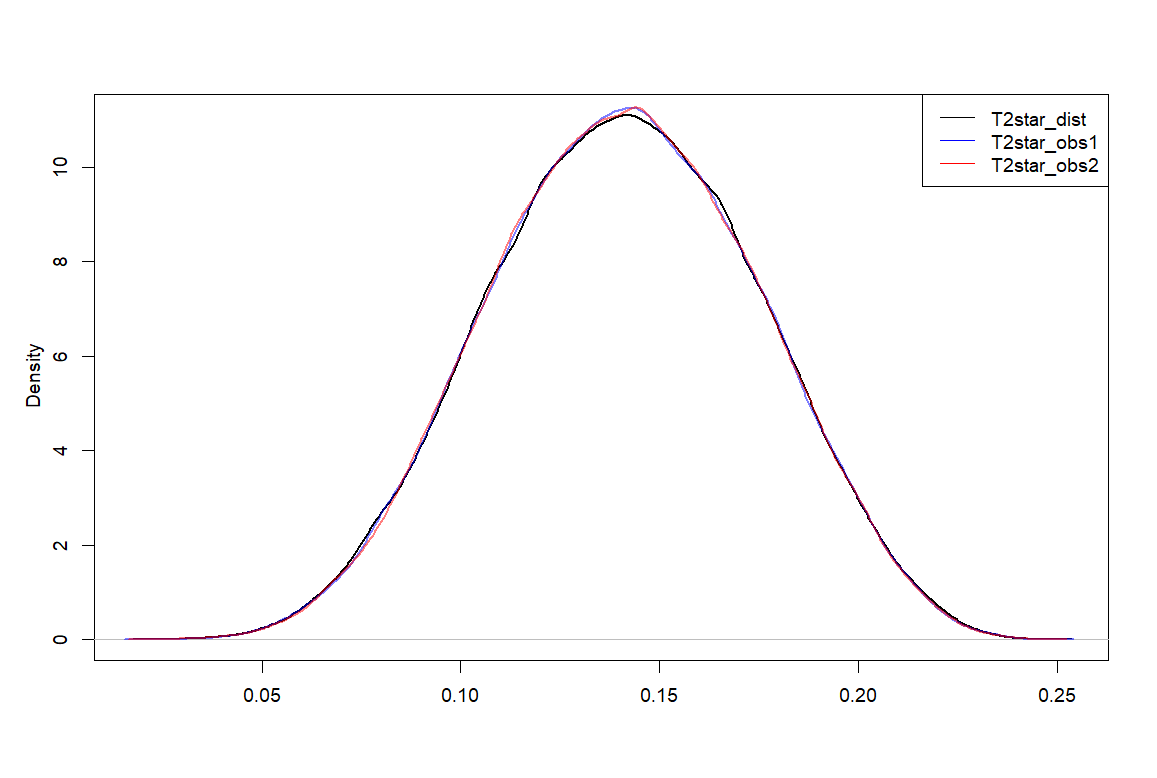}
  \caption{$n=10$}
  \label{fig:Sph10}
\end{subfigure}
\begin{subfigure}{0.48\textwidth}
  \centering
  \includegraphics[width=\linewidth]{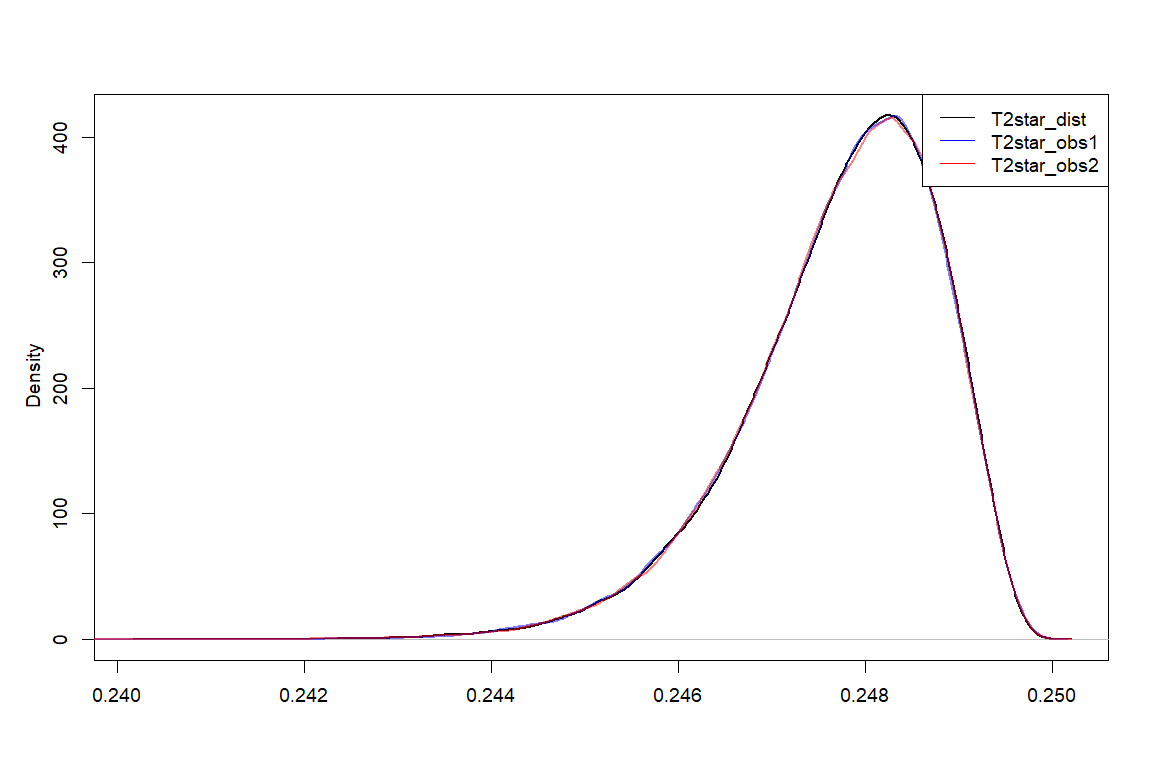}
  \caption{$n=500$}
  \label{fig:Sph500}
\end{subfigure}
\caption{Graph for the Sphericity Test statistic's distribution along with the empirical distribution of the observed values of $T_2^\star$ for given $n=10, 500$, $p=4$, \(\vmu\), \(\mSigma_1\) (T2star\_obs1) and  \(\mSigma_2\) (T2star\_obs2).}
\label{fig:Sph}
\end{figure}

\begin{figure}[!ht]
\begin{subfigure}{0.48\textwidth}
  \centering
  \includegraphics[width=\linewidth]{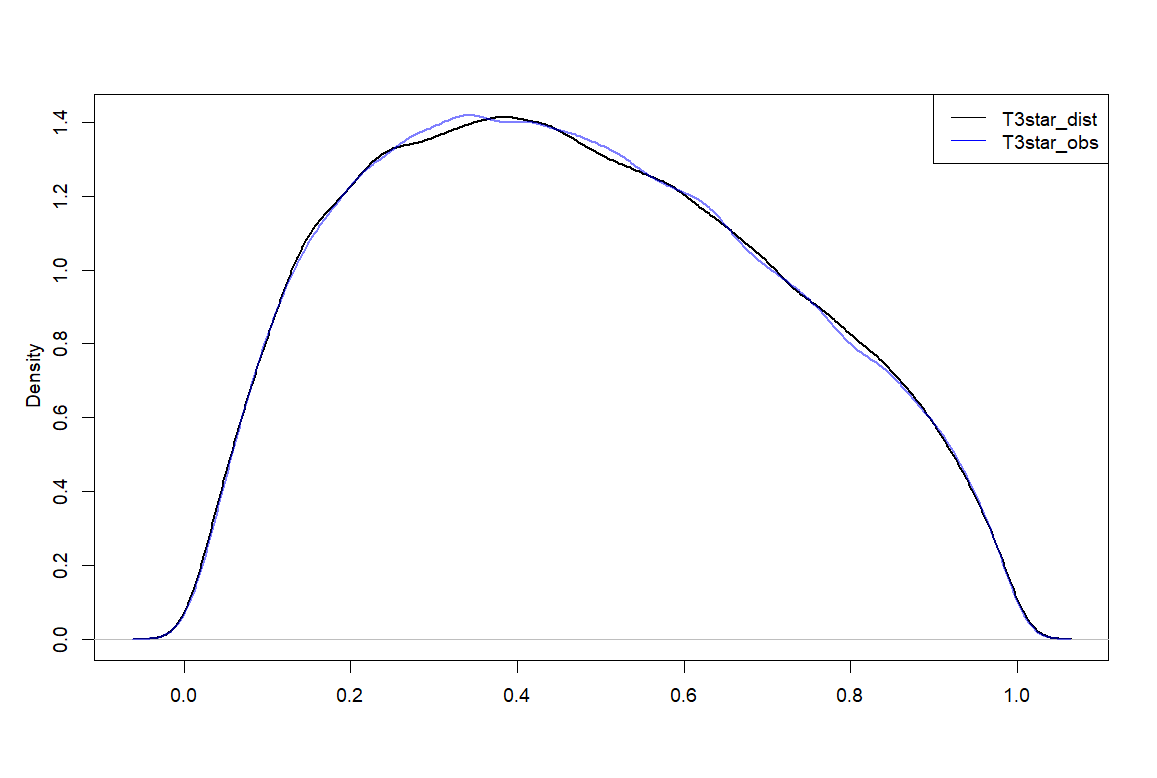}
  \caption{$p_1=1$ and $\mSigma=\mSigma_1$}
  \label{fig:Ind_1_10}
\end{subfigure}
\begin{subfigure}{0.48\textwidth}
  \centering
  \includegraphics[width=\linewidth]{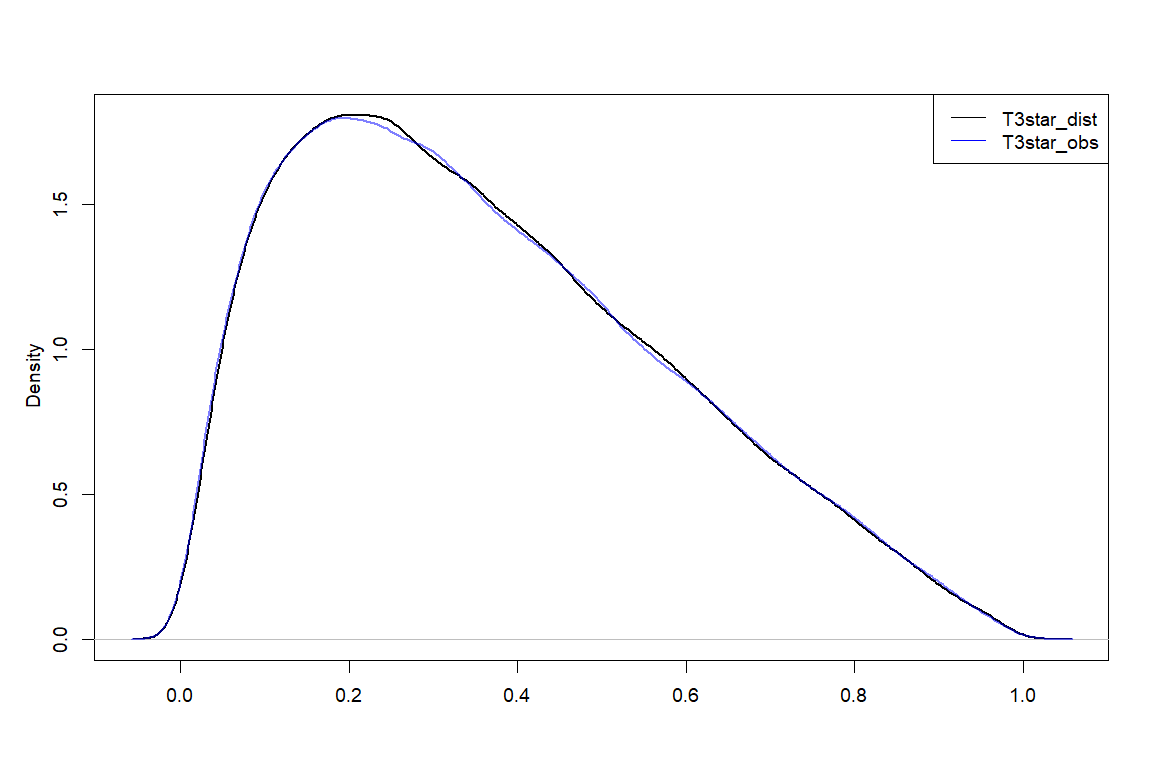}
  \caption{$p_1=2$ and $\mSigma=\mSigma_4$}
  \label{fig:Ind_2_10}
\end{subfigure}
\caption{Graph for the Independence Test statistic's distribution along with the empirical distribution of the observed values of $T_3^\star$ for given $n=10$, $p=4$, \(\vmu\), \(\mSigma_1\) (T3star\_obs1) and  \(\mSigma_4\) (T3star\_obs2).}
\label{fig:Ind1}
\end{figure}

\begin{figure}[!ht]
\begin{subfigure}{0.45\textwidth}
  \centering
  \includegraphics[width=\linewidth]{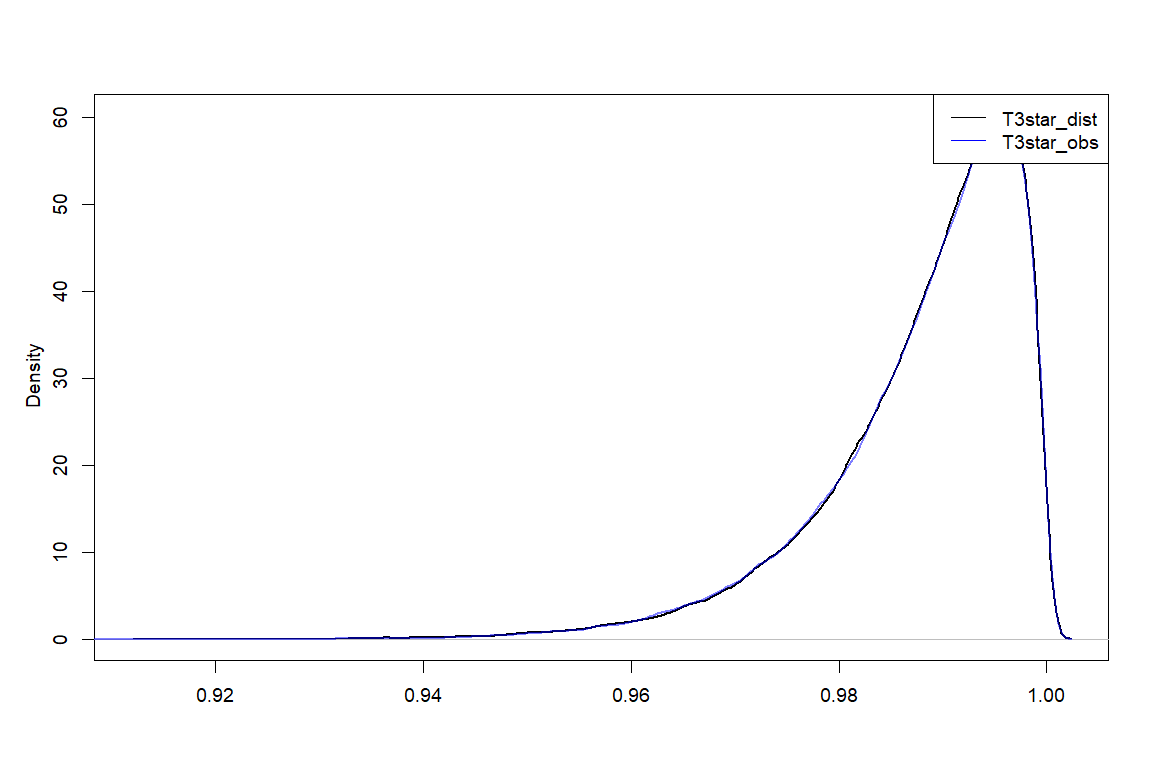}
  \caption{$p_1=1$ and $\mSigma=\mSigma_1$}
  \label{fig:Ind_1_500}
\end{subfigure}
\begin{subfigure}{0.45\textwidth}
  \centering
  \includegraphics[width=\linewidth]{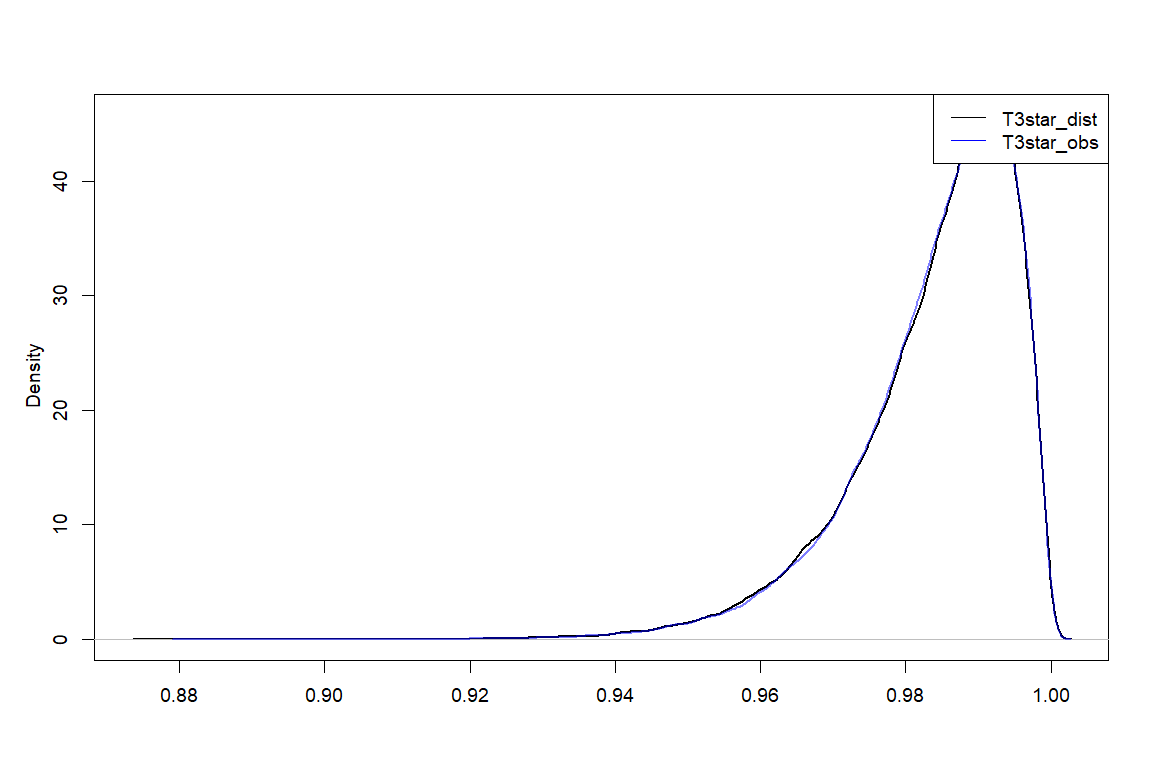}
  \caption{$p_1=2$ and $\mSigma=\mSigma_4$}
  \label{fig:Ind_2_500}
\end{subfigure}
\caption{Graph for the Independence Test statistic's distribution along with the empirical distribution of the observed values of $T_3^\star$ for given $n=500$, $p=4$, \(\vmu\), \(\mSigma_1\) (T3star\_obs1) and  \(\mSigma_4\) (T3star\_obs2).}
\label{fig:Ind2}
\end{figure}

\begin{figure}[!ht]
\begin{subfigure}{0.45\textwidth}
  \centering
  \includegraphics[width=\linewidth]{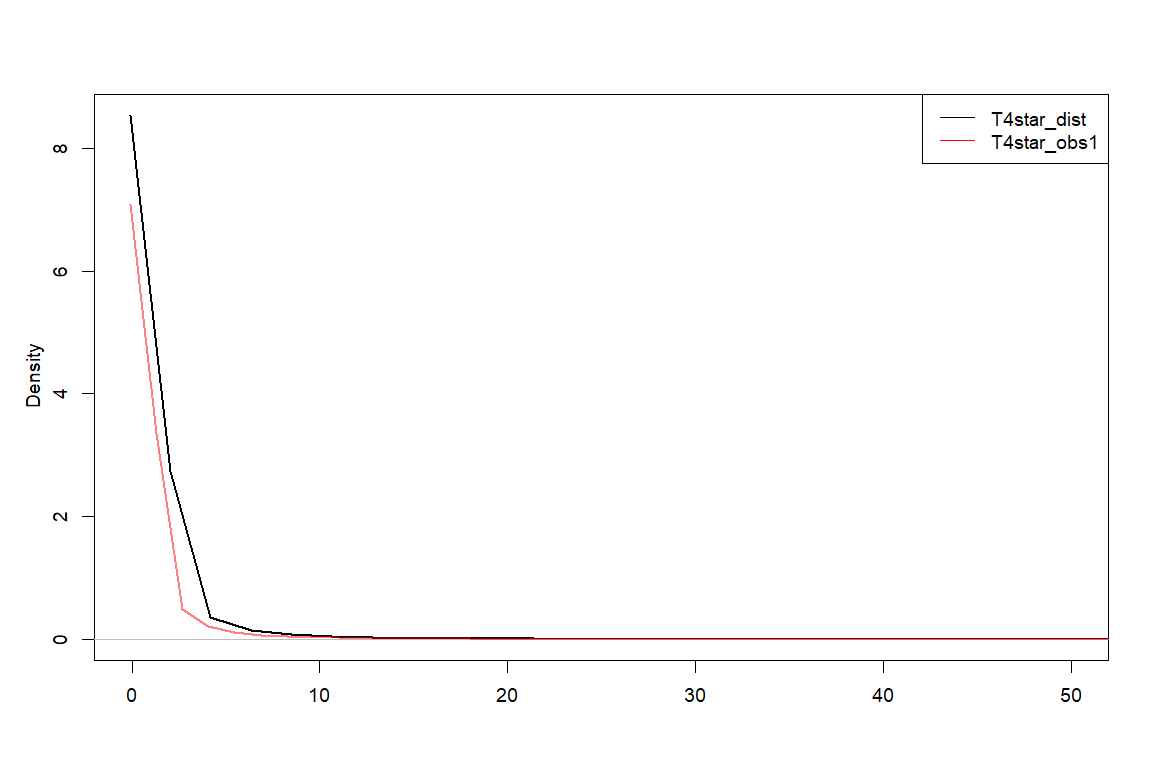}
  \caption{$p_1=2$ and $\mSigma=\mSigma_3$}
  \label{fig:Reg_1_10}
\end{subfigure}
\begin{subfigure}{0.45\textwidth}
  \centering
  \includegraphics[width=\linewidth]{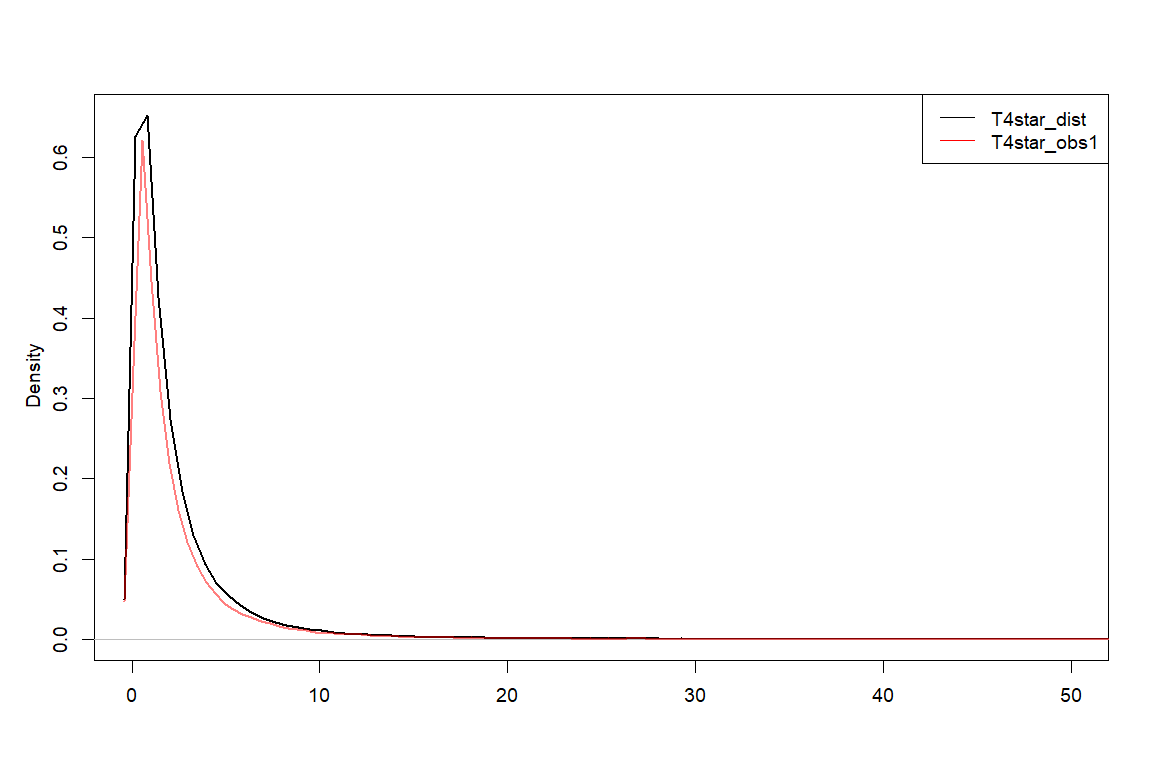}
  \caption{$p_1=1$ and $\mSigma=\mSigma_4$}
  \label{fig:Reg_2_10}
\end{subfigure}
\caption{Graph for the Regression Test statistic's distribution along with the empirical distribution of the observed values of $T_4^\star$ for given $n=10$, $p=4$, \(\vmu\), \(\mSigma_3\) (T4star\_obs1) and  \(\mSigma_4\) (T4star\_obs2).}
\label{fig:Reg1}
\end{figure}

\begin{figure}[!ht]
\begin{subfigure}{0.47\textwidth}
  \centering
  \includegraphics[width=\linewidth]{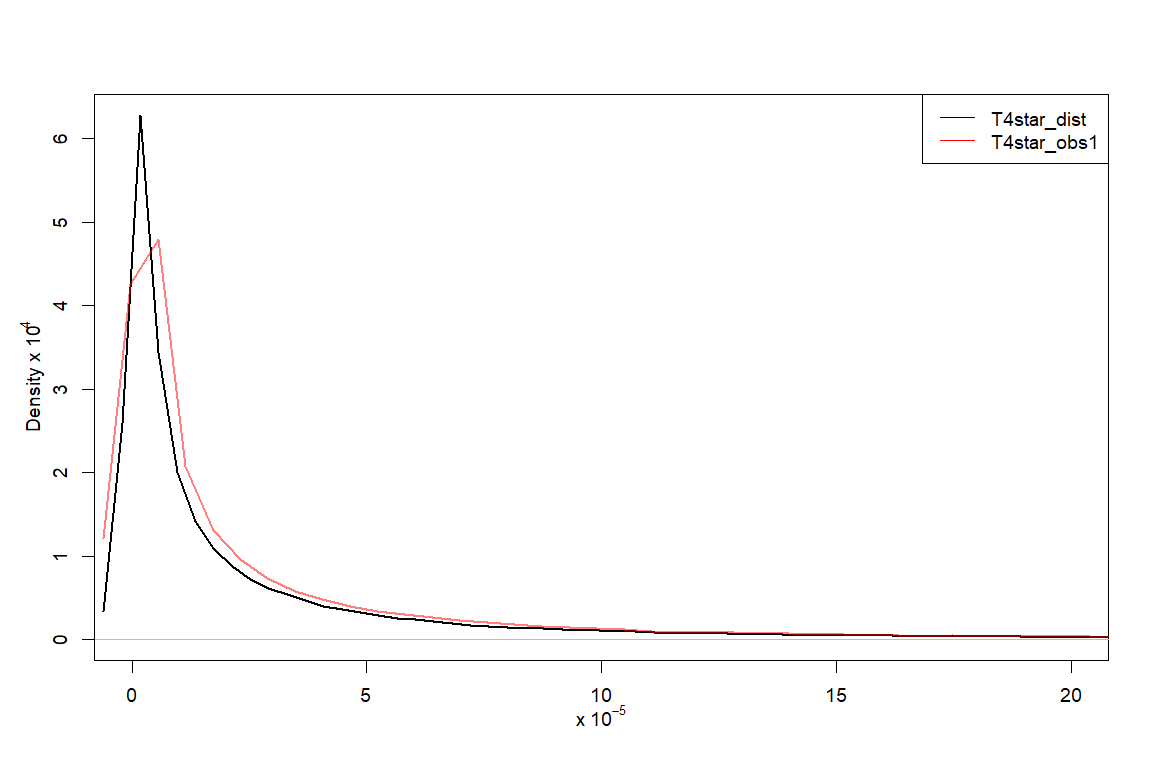}
  \caption{$p_1=2$ and $\mSigma=\mSigma_3$}
  \label{fig:Reg_1_500}
\end{subfigure}
\begin{subfigure}{0.47\textwidth}
  \centering
  \includegraphics[width=\linewidth]{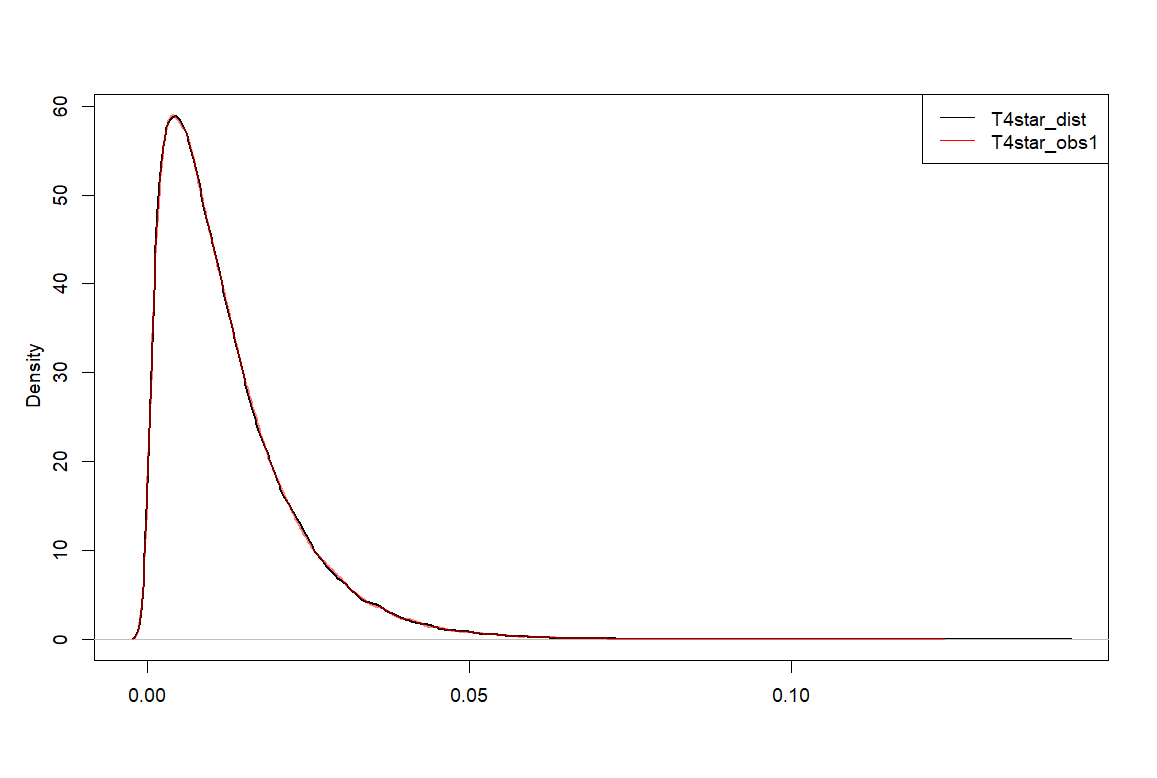}
  \caption{$p_1=1$ and $\mSigma=\mSigma_4$}
  \label{fig:Reg_2_500}
\end{subfigure}
\caption{Graph for the Regression Test statistic's distribution along with the empirical distribution of the observed values of $T_4^\star$ for given $n=500$, $p=4$, \(\vmu\), \(\mSigma_3\) (T4star\_obs1) and  \(\mSigma_4\) (T4star\_obs2).}
\label{fig:Reg2}
\end{figure}

From Table \ref{tab:cov}, we observe that all values of \(cov\) are approximately equal to the nominal value of 0.95, as expected, due to the exact nature of the inference procedures across all tests and under different conditions. Furthermore, to better illustrate the behavior of the random variables discussed in Section \ref{sec:Notation}, Figures \ref{fig:GV}, \ref{fig:Sph}, \ref{fig:Ind1}, \ref{fig:Ind2}, \ref{fig:Reg1}, and \ref{fig:Reg2}  display the empirical distributions of the observed values from the synthetic datasets, generated under varying simulation conditions. To simplify we only include the distributions for \(n=10\) and \(n=500\). These empirical distributions are presented alongside the corresponding theoretical distributions created using the functions \texttt{GVdist}, \texttt{Sphdist}, \texttt{Inddist}, and \texttt{Canodist}, allowing for the comparison and the validation of the theoretical results.

\section{Summary}\label{sec:5}

In this article, we introduced \texttt {PSInference}, a package aimed to generate fully synthetic datasets via Plug-in Sampling method and perform exact inferential procedures based on multivariate normal data. The package focuses on cases where only a single synthetic dataset is available, which is common in scenarios where multiple imputations are prohibitive. Nevertheless, the authors plan to extend the inferential procedures for the multiple imputation scenario.

We illustrated the core functions of \verb|PSInference|, including the generation of synthetic datasets using \texttt{simSynthData} function, and the functions for inferential procedures such as computing confidence intervals for the generalized variance (\texttt{GVdist}), tests for the sphericity (\texttt{Sphdist}), the independence between two subsets of variables (\texttt{Inddist}), and the regression of one set of variables on another (\texttt{Canodist}).

Our numerical studies, conducted through Monte Carlo simulations, verified the accuracy of the implemented procedures by estimating the coverage probabilities (\(cov\)) for different sample sizes and different covariance matrix structures. These results confirmed that the coverage probabilities closely match the nominal value of 0.95 across all tests, validating the robustness of the exact inferential procedures provided by the package.

Furthermore, we presented empirical distributions of the test statistics for each inferential procedure, computed from the synthetic datasets generated during the simulations. These empirical distributions were compared to their theoretical counterparts.

\section*{R Software}
The R package \href{https://cran.r-project.org/web/packages/PSinference/index.html}{PSInference} is now available on the CRAN website (\url{https://cran.r-project.org/web/packages/PSinference/index.html})

\section*{Acknowledgments}
The work of authors is funded by national funds through the FCT – Funda\c{c}\H{a}o para a Ci\^{e}ncia e a Tecnologia, I.P., under the scope of the projects UIDB/00297/2020 (\url{https://doi.org/10.54499/UIDB/00297/2020}) and UIDP/00297/2020 (\url{https://doi.org/10.54499/UIDP/00297/2020}) (Center for Mathematics and Applications)”.

\end{document}